# The two mode Heaviside equation for ultra-short laser pulses interaction with matter

J. Marciak – Kozłowska

Institute of Electron Technology,

Al. Lotników 32/46, 02 – 668 Warsaw,    Poland

and

M .Pelc ,  M.  Kozlowski

**Abstract**

In his paper the two mode Heaviside equation was formulated and solved. It was shown that the interaction of ultra-short laser pulses with matter leads to two mode excitation electrons and phonons which afterwards diffuse with finite velocities.

# 1. INTRODUCTION

Dynamics of nonequilibrium electrons and phonons in metals, semiconductors have been the focus of much attention because of their fundamental interest in solid state physics and nanotechnology.

In metals, relaxation dynamics of optically excited nonequilibrium electrons has been extensively studied by pump – probe techniques using femtosecond lasers [1 – 4].

Recently [5] it was shown that the optically excited metals relax to equilibrium with two models: rapid electron relaxation and slow thermal relaxation through the creation of the optical phonons.

In this paper we develop the hyperbolic thermal diffusion equation with two models: electrons and phonons relaxation. These two modes are characterized by two relaxation times $\tau_1$ for electrons and $\tau_2$ for phonons. This new equation is the generalization of our one mode hyperbolic equation, with only electrons degrees of freedom, $\tau_1$ [ 6 ]. The hyperbolic two mode equation is the analogous equation to Klein – Gordon equation and allows the heat propagation with finite speed.

# 2. THE MODEL

As was shown in paper [6] for high frequency laser pulses the diffusion velocity exceeds that of light. This is not possible and merely demonstrates that Fourier equation is not really correct. Oliver Heaviside was well aware of this writing [7] :

*All diffusion formulae (as in heat conduction) show instantaneous action to an infinite distances of a source, though only to an infinitesimal extent. To make the theory of heat diffusion be rational as well as practical some modification of the equation to remove the instantaneity, however little difference it may make quantatively, in general.*

August 1876 saw the appearance in Philosophical Magazine [ 7 ] the paper which extended the mathematical understanding the diffusion (Fourier) equation. O. Heaviside for the first time wrote the hyperbolic diffusion equation for the voltage *V(x,t)*. assuming a uniform resistance, capacitance and inductance per unit length, *k, c* and *s* respectively he arrived at:

$$\frac{\partial^2 V(x,t)}{\partial x^2} = kc\frac{\partial V(x,t)}{\partial t} + sc\frac{\partial^2 V(x,t)}{\partial t^2} \tag{1}$$

The discussion of the broad sense of the Heaviside equation (1) can be find out, for example in our monograph [ 8], viz,

$$\tau^2 \frac{\partial^2 T}{\partial t^2} + \tau\frac{\partial T}{\partial t} + \frac{2V\tau}{\hbar}T = \tau\frac{\hbar}{m}\nabla^2 T \tag{2}$$

In Eq. (2) $T(\vec{r},t)$ denotes the temperature field, *V* is the external potential, *m* is the mass of heat carrier and $\tau$ is the relaxation time

$$\tau = \frac{\hbar}{mv^2} \tag{3}$$

As can be seen from formulae (2) and (3) in hyperbolic diffusion equation the same relaxation time $\tau$ is assumed for both type of motion: wave and diffusion.

This can not be so obvious. For example let us consider the simpler harmonic oscillator equation:

$$m\frac{d^2x}{dt^2} + kx + c\frac{dx}{dt} = 0 \tag{4}$$

Equation (4) can be written as

$$\tau^2\frac{d^2x}{dt^2} + \tau\frac{dx}{dt} + x = 0 \tag{5}$$

where

$$\tau^2 = \frac{m}{k}, \qquad \tau = \frac{c}{k} \tag{6}$$

i.e.

$$c^2 = km \tag{7}$$

As it was well known equation (5) with formula (6) describes only the weakly damped (periodic) motion of the harmonic oscillator (HO). It must be stressed that for HO exists also critically damped and overdamped modes which are not describes by the equation (5). The general master equation for HO must be written as

$$\tau_1^2 \frac{d^2 x}{dt^2} + \tau_2 \frac{dx}{dt} + x = 0 \qquad (8)$$

Following the discussion of the formulae (5) to (8) we argue that the general hyperbolic diffusion equation can be written as:

$$\tau_1^2 \frac{\partial^2 T}{\partial t^2} + \tau_2 \frac{\partial T}{\partial t} + \frac{2V\tau_2}{\hbar} T = \frac{\hbar}{m} \tau_2 \nabla^2 T \qquad (9)$$

and $\tau_1 \neq \tau_2$

Equation (9) describes the temperature field generated by ultra-short laser pulses. In Eq. (9) two modes: wave and diffusion are described by different relaxation times.

For quantum hyperbolic equation (9) we seek solution in the form (in 1D)

$$T(x,t) = e^{-\frac{t\tau_2}{2\tau_1^2}} u(x,t) \qquad (10)$$

After substitution Eq. (10) into Eq. (9) one obtains

$$\frac{1}{v^2} \frac{\partial^2 u}{\partial t^2} - \frac{\partial^2 u}{\partial x^2} + qu(x,t) = 0 \qquad (11)$$

where

$$v^2 = \frac{\hbar \tau_2}{m \tau_1^2}, \qquad q = \left( \frac{2Vm}{\hbar^2} - \frac{1}{4} \frac{m}{\hbar} \frac{\tau_2}{\tau_1^2} \right) \qquad (12)$$

Equation (11) is the thermal two-mode Klein – Gordon equation and is the generalization of Klein – Gordon one mode equation developed in our monograph.

For Cauchy initial condition

$$u(x,0) = f(x), \qquad \frac{\partial u(x,0)}{\partial t} = g(x) \qquad (13)$$

the solution of Eq. (11) has the form

$$u(x,t) = \frac{f(x-vt) + f(x+vt)}{2}$$
$$+ \frac{1}{2v} \int_{x-vt}^{x+vt} g(\varsigma) I_0 \left[ \sqrt{-q(v^2 t^2 - (x-\varsigma)^2)} \right] d\varsigma \qquad (14)$$
$$+ \frac{(v\sqrt{-q})t}{2} \int_{x-vt}^{x+vt} f(\varsigma) \frac{I_1 \left[ \sqrt{-q(v^2 t^2 - (x-\varsigma)^2)} \right]}{\sqrt{v^2 t^2 - (x-\varsigma)^2}} d\varsigma$$

for *q* < 0

and

$$u(x,t) = \frac{f(x-vt) + f(x+vt)}{2} + \frac{1}{2v}\int_{x-vt}^{x+vt} g(\varsigma) J_0\left[\sqrt{q(v^2t^2-(x-\varsigma)^2)}\right]d\varsigma - \frac{(v\sqrt{q})t}{2}\int_{x-vt}^{x+vt} f(\varsigma) \frac{J_1\left[\sqrt{q(v^2t^2-(x-\varsigma)^2)}\right]}{\sqrt{v^2t^2-(x-\varsigma)^2}}d\varsigma \quad (15)$$

for *q* > 0.

## 3. CONCLUSIONS

In this paper the new thermal Klein -Gordon equation for laser beam interaction with matter was developed and solved. It was argued that in the heating of the matter with laser pulses two modes of the excitation can be observed : fast thermal wave and slow diffusion.